\documentclass[runningheads]{llncs}
\usepackage{graphicx}
\usepackage{comment}
\usepackage{amsmath,amssymb} 
\usepackage{subfigure}
\usepackage{array}
\usepackage{booktabs}
\usepackage{colortbl}
\usepackage{hhline}
\usepackage{arydshln}
\usepackage{verbatim} 
\usepackage{gensymb} 
\usepackage{multirow}
\usepackage{tabu}

\usepackage{times}
\usepackage{epsfig}
\usepackage{caption}
\usepackage{ulem}

\newcommand\blfootnote[1]{%
	\begingroup
	\renewcommand\thefootnote{}\footnote{#1}%
	\addtocounter{footnote}{-1}%
	\endgroup
}

\usepackage[table]{xcolor}
\definecolor{lightgray}{gray}{0.9}

\graphicspath{{IMAGES/}}

\usepackage[pagebackref=true,breaklinks=true,letterpaper=true,colorlinks,bookmarks=false]{hyperref}
\usepackage{imakeidx}
\makeindex

\begin{document}
\pagestyle{headings}
\mainmatter
\def\ECCVSubNumber{2629}  

\title{Stochastic Frequency Masking to Improve Super-Resolution and Denoising Networks} 

\titlerunning{SFM}
\index{El Helou, Majed}
\author{Majed El Helou\inst{*} \and
Ruofan Zhou\inst{*} \and
Sabine S\"usstrunk}
%
\institute{School of Computer and Communication Sciences, EPFL, Switzerland
\email{\{majed.elhelou,ruofan.zhou,sabine.susstrunk\}@epfl.ch}}
\maketitle

\begin{abstract}
Super-resolution and denoising are ill-posed yet fundamental image restoration tasks. In blind settings, the degradation kernel or the noise level are unknown. This makes restoration even more challenging, notably for learning-based methods, as they tend to overfit to the degradation seen during training. 
We present an analysis, in the frequency domain, of degradation-kernel overfitting in super-resolution and introduce a conditional learning perspective that extends to both super-resolution and denoising. Building on our formulation, we propose a stochastic frequency masking of images used in training to regularize the networks and address the overfitting problem. Our technique improves state-of-the-art methods on blind super-resolution with different synthetic kernels, real super-resolution, blind Gaussian denoising, and real-image denoising.
\end{abstract}

\section{Introduction} \label{sec:introduction}
Image super-resolution (SR) and denoising are fundamental restoration tasks widely applied in imaging pipelines. They are crucial in various applications, such as medical imaging~\cite{li2012group,peled2001superresolution,shi2013cardiac}, low-light imaging~\cite{chatterjee2011noise}, astronomy~\cite{beckouche2013astronomical}, satellite imaging~\cite{benazza2005building,thornton2006sub}, or face detection~\cite{gunturk2003eigenface}. However, both are challenging ill-posed inverse problems. \blfootnote{$^*$ The first two authors have similar contributions.}

Recent learning methods based on convolutional neural networks (CNNs) achieve better restoration performance than classical approaches, both in SR and denoising. CNNs are trained on large datasets, sometimes real~\cite{KMSR} but often synthetically generated with either one kernel or a limited set~\cite{RRDB,RCAN}. They learn to predict the restored image or the residual between the restored target and the input~\cite{kim2016accurate,dncnn}. However, to be useful in practice, the networks should perform well on test images with unknown degradation kernels for SR, and unknown noise levels for denoising. Currently, they tend to overfit to the set of degradation models seen during training\textcolor{blue}{~\cite{efrat2013accurate}}.

\begin{figure}[t]
	\centering
	\includegraphics[width=.7\linewidth,trim={32 0 70 18},clip]{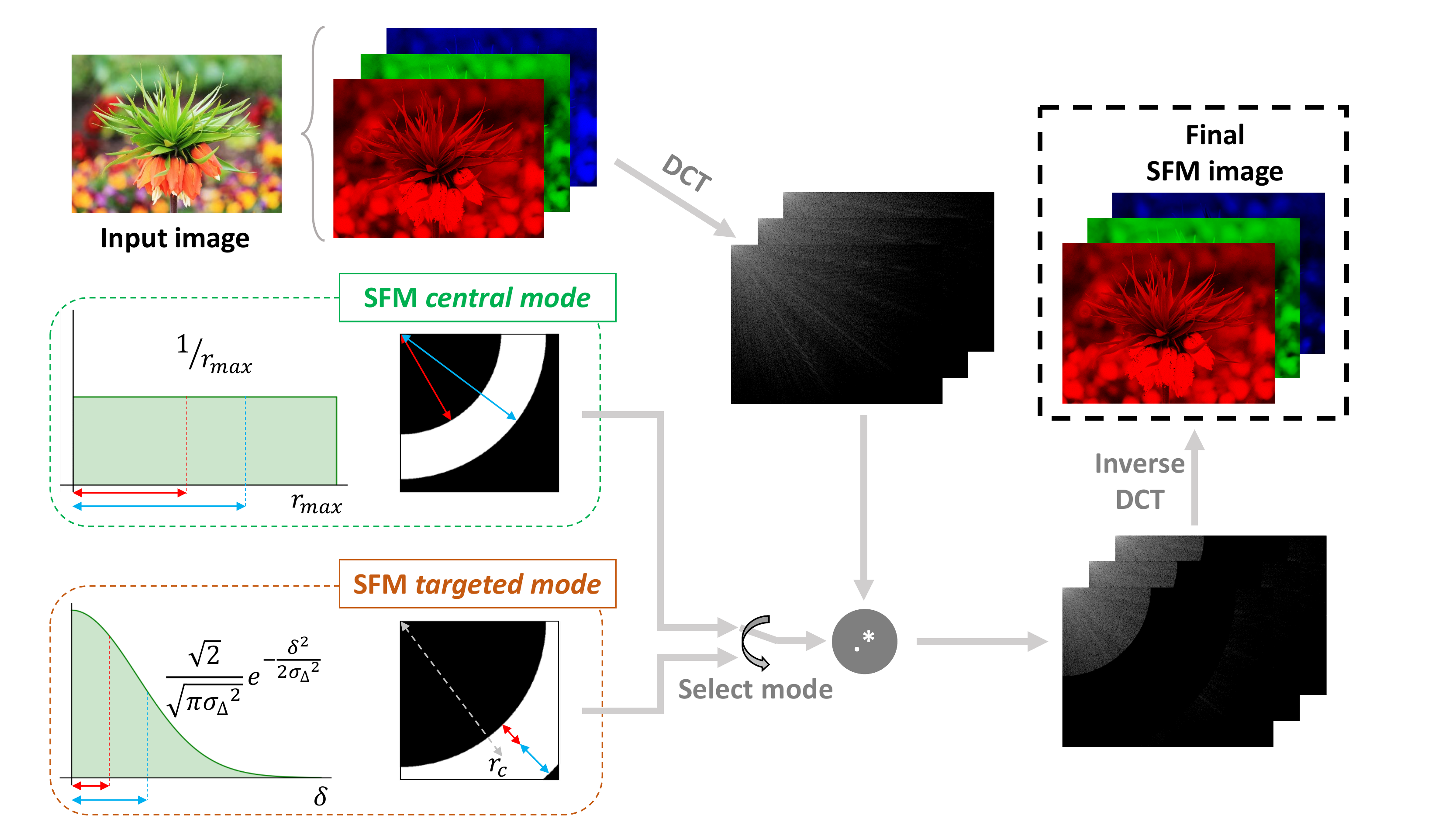}
	\caption{Overview of our Stochastic Frequency Masking (SFM). In the \textit{central mode}, two radii values are sampled uniformly to delimit a masking area, and in the \textit{targeted mode}, the sampled values delimit a quarter-annulus away from a target frequency. The obtained mask, visualized with inverted color, is applied channel-wise to the discrete cosine transform (DCT) of the input image. Inverting back to the spatial domain yields the final SFM image, which we integrate into the training of SR and denoising networks.}
	\label{fig:teaser}
\end{figure}

We investigate the SR degradation-kernel overfitting with an analysis in the frequency domain. Our analysis reveals that an implicit conditional learning is taking place in SR networks, namely, the learning of residual high-frequency content given low frequencies. We additionally show that this result extends to denoising as well. Building on our insights, we present \textbf{S}tochastic \textbf{F}requency \textbf{M}asking (SFM), which stochastically masks frequency components of the \textit{images used in training}. Our SFM method (Fig.~\ref{fig:teaser}) is applied to a subset of the training images to regularize the network. It encourages the conditional learning to improve SR and denoising networks, notably when training under the challenging blind conditions. It can be applied during the training of \textit{any} learning method, and has no additional cost at test time. \blfootnote{Code available at: \url{https://github.com/majedelhelou/SFM}}

Our experimental results show that SFM improves the restoration performance of state-of-the-art networks on blind SR tasks as well as on blind denoising tasks. For blind SR, we conduct experiments on synthetic bicubic and Gaussian degradation kernels, and on real degraded images. For blind denoising, we conduct experiments on additive white Gaussian denoising and on real microscopy Poisson-Gaussian image denoising. SFM improves the performance of state-of-the-art networks on each of these tasks.

Our contributions are summarized as follows. We present a frequency-domain analysis of the degradation-kernel overfitting of SR networks, and highlight the implicit conditional learning that, as we also show, extends to denoising. 
We present a novel technique, SFM, that regularizes the learning of SR and denoising networks by only filtering the training data. It allows the networks to better restore frequency components and avoid overfitting. 
We empirically show that SFM improves the results of state-of-the-art learning methods on blind SR with different synthetic degradations, real-image SR, blind Gaussian denoising, and real-image denoising on high noise levels.

\section{Related work} \label{sec:related_work}
\textbf{Super-resolution}. Depending on their image prior, SR algorithms can be divided into prediction models~\cite{schulter2015fast}, edge-based models~\cite{chan2009neighbor}, gradient-profile pior methods~\cite{sun2008image} and example-based methods~\cite{freeman2002example}. Deep example-based SR networks hold the state-of-the-art performance. 
Zhang~\textit{et al.} propose a very deep architecture based on residual channel attention to further improve these networks~\cite{RCAN}. It is also possible to train in the wavelet domain to improve the memory and time efficiency of the networks~\cite{zhou2019comparative}. 
Perceptual loss~\cite{johnson2016perceptual} and GANs~\cite{ledig2017photo,RRDB} are leveraged to mitigate blur and push the SR networks to produce more visually-pleasing results. However, these networks are trained using a limited set of kernels, and studies have shown that they have poor generalization to unseen degradation kernels~\cite{IKC,ZSSR}.
To address blind SR, which is degradation-agnostic, recent methods propose to incorporate the degradation parameters including the blur kernel into the network~\cite{ZSSR,IRCNN,SRMD,DPP}. However, these methods rely on blur-kernel estimation algorithms and thus have a limited ability to handle arbitrary blur kernels. 
The most recent methods, namely IKC~\cite{IKC} and KMSR~\cite{KMSR}, propose kernel estimation and modeling in their SR pipeline. However, it is hard to gather enough training kernels to cover the real-kernel manifold, while also ensuring effective learning and avoiding that these networks overfit to the chosen kernels.
Recently, real-image datasets were proposed~\cite{realsrdata,zoomindata} to enable SR networks to be trained and tested on high- and low-resolution (HR-LR) pairs, which capture the same scene but at different focal lengths. These datasets are also limited to the degradations of only a few cameras and cannot guarantee that SR models trained on them would generalize to unseen degradations.
Our SFM method, which builds on our degradation-kernel overfitting analysis and our conditional learning perspective, can be used to improve the performance of \textit{all} the SR networks we evaluate, including ones that estimate and model degradation kernels.

\textbf{Denoising}. Classical denoisers such as PURE-LET~\cite{luisier2011image}, which is specifically aimed at Poisson-Gaussian denoising, KSVD~\cite{aharon2006k}, WNNM~\cite{gu2014weighted}, BM3D~\cite{BM3D}, and EPLL~\cite{zoran2011learning}, which are designed for Gaussian denoising, have the limitation that the noise level needs to be known at test time, or at least estimated~\cite{foi2008practical}.
Recent learning-based denoisers outperform the classical ones on Gaussian denoising~\cite{anwar2019real,plotz2018neural,dncnn}, but require the noise level~\cite{ffdnet}, or pre-train multiple models for different noise levels~\cite{lefkimmiatis2018universal,IRCNN}, or more recently attempt to predict the noise level internally~\cite{el2020blind}. For a model to work under blind settings and adapt to any noise level, a common approach is to train the denoiser network while varying the training noise level~\cite{anwar2019real,plotz2018neural,dncnn}. 
Other recent methods, aimed at real-image denoising such as microscopy imaging~\cite{zhang2019poisson}, learn image statistics without requiring ground-truth samples on which noise is synthesized.
This is practical because ground-truth data can be extremely difficult and costly to acquire in, for instance, medical applications. Noise2Noise~\cite{lehtinen2018noise2noise} learns to denoise from pairs of noisy images. The noise is assumed to be zero in expectation and decorrelated from the signal. Therefore, unless the network memorizes it, the noise would not be predicted by it, and thus gets removed~\cite{lehtinen2018noise2noise,ulyanov2018deep}. Noise2Self~\cite{batson2019noise2self}, which is a similar but more general version of Noise2Void~\cite{krull2019noise2void}, also assumes the noise to be decorrelated, conditioned on the signal. The network learns from single noisy images, by learning to predict an image subset from a separate subset, again with the assumption that the noise is zero in expectation. Although promising, these two methods do not yet reach the performance of Noise2Noise. By regularizing the conditional learning defined from our frequency-domain perspective, our SFM method improves the high noise level results of \textit{all} tested denoising networks, notably under blind settings.

One example that uses frequency bands in restoration is the method in~\cite{anwar2018image} that defines a prior based on a distance metric between a test image and a dataset of same-class images used for a deblurring optimization.
The distance metric computes differences between image frequency bands. In contrast, we apply frequency masking on training images to regularize deep learning restoration networks, improving performance and generalization.  
Spectral dropout~\cite{khan2019regularization} regularizes network activations by dropping out components in the frequency domain to remove the least relevant, while SFM regularizes training by promoting the conditional prediction of different frequency components through masking the training images themselves.
The most related work to ours is a recent method proposed in the field of speech recognition~\cite{park2019specaugment}. The authors augment speech data in three ways, one of which is in the frequency domain. It is a random separation of frequency bands, which splits different speech components to allow the network to learn them one by one. A clear distinction with our approach is that we do not aim to separate input components to be each individually learned. Rather, we mask targeted frequencies from the \textit{training} input to strengthen the conditional frequency learning, and indirectly simulate the effect of a variety of kernels in SR and noise levels in denoising. The method we present is, to the best of our knowledge, the first frequency-based input \textit{masking} method to regularize SR and denoising training.


\section{Frequency perspective on SR and denoising} \label{sec:Freq_perspective}
\subsection{Super-resolution} \label{subsec:SFM_SR}
\subsubsection{Preliminaries} \label{subsubsec:SFM_SR_math}
Downsampling, a key element in modeling SR degradation, can be well explained in the frequency domain where it is represented by the sum of shifted and stretched versions of the frequency spectrum of a signal. Let $x$ be a one-dimensional discrete signal, e.g., a pixel row in an image, and let $z$ be a downsampled version of $x$ with a sampling interval $T$. In the discrete-time Fourier transform domain, with frequencies $\omega \in [-\pi,\pi]$, the relation between the transforms $X$ and $Z$ of the signals $x$ and $z$, respectively, is given by $Z(\omega) = \frac{1}{T} \sum_{k=0}^{T-1} X((\omega+2\pi k)/T)$. The $T$ replicas of $X$ can overlap in the high frequencies and cause aliasing. Aside from complicating the inverse problem of restoring $x$ from $z$, aliasing can create visual distortions. Before downsampling, low-pass filtering is therefore applied to attenuate if not completely remove the high-frequency components that would otherwise overlap.

These low-pass filtering blur kernels are applied through a spatial convolution over the image. The set of real kernels spans only a subspace of all mathematically-possible kernels. This subspace is, however, not well-defined analytically and, in the literature, is often limited to the non-comprehensive subspace spanned by 2D Gaussian kernels. Many SR methods thus model the anti-aliasing filter as a 2D Gaussian kernel, attempting to mimic the point spread function (PSF) of capturing devices~\cite{dong2012nonlocally,shi2016real,yang2014single}. 
In practice, even a single imaging device results in multiple kernels, depending on its settings~\cite{el2018aam}.
For real images, the kernel can also be different from a Gaussian kernel~\cite{efrat2013accurate,IKC}.  
The essential point is that the anti-aliasing filter causes the loss of high-frequency components, and that this filter can differ from image to image.

\begin{figure}[t!]
	\centering
	\subfigure[Experimental setup]{\includegraphics[width=.8\linewidth,trim={0 0 5 20},clip]{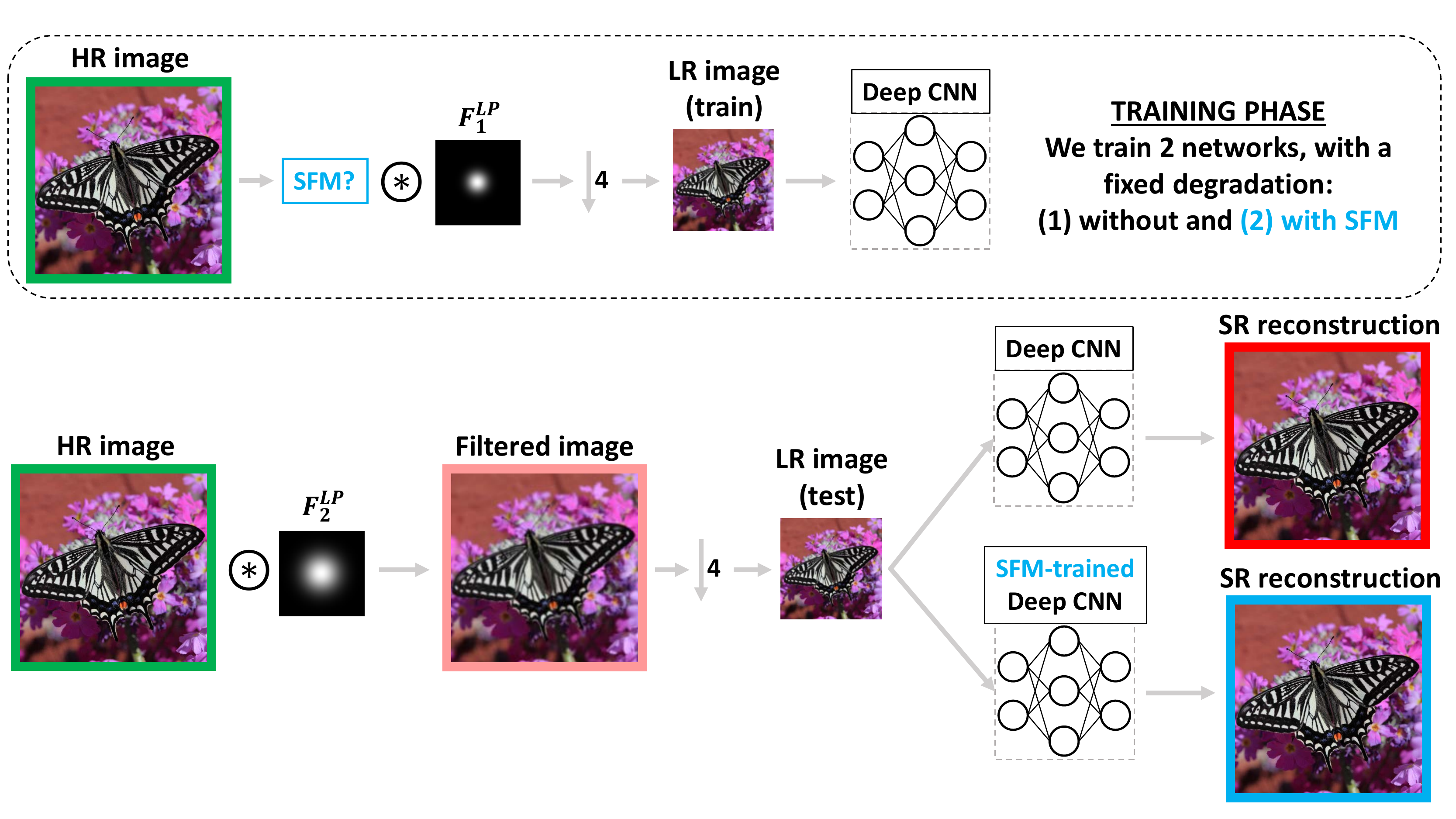}
		\label{fig:pipeline_hole}}
	\subfigure[Without SFM]{\includegraphics[width=.49\linewidth,trim={7 6 8 5},clip]{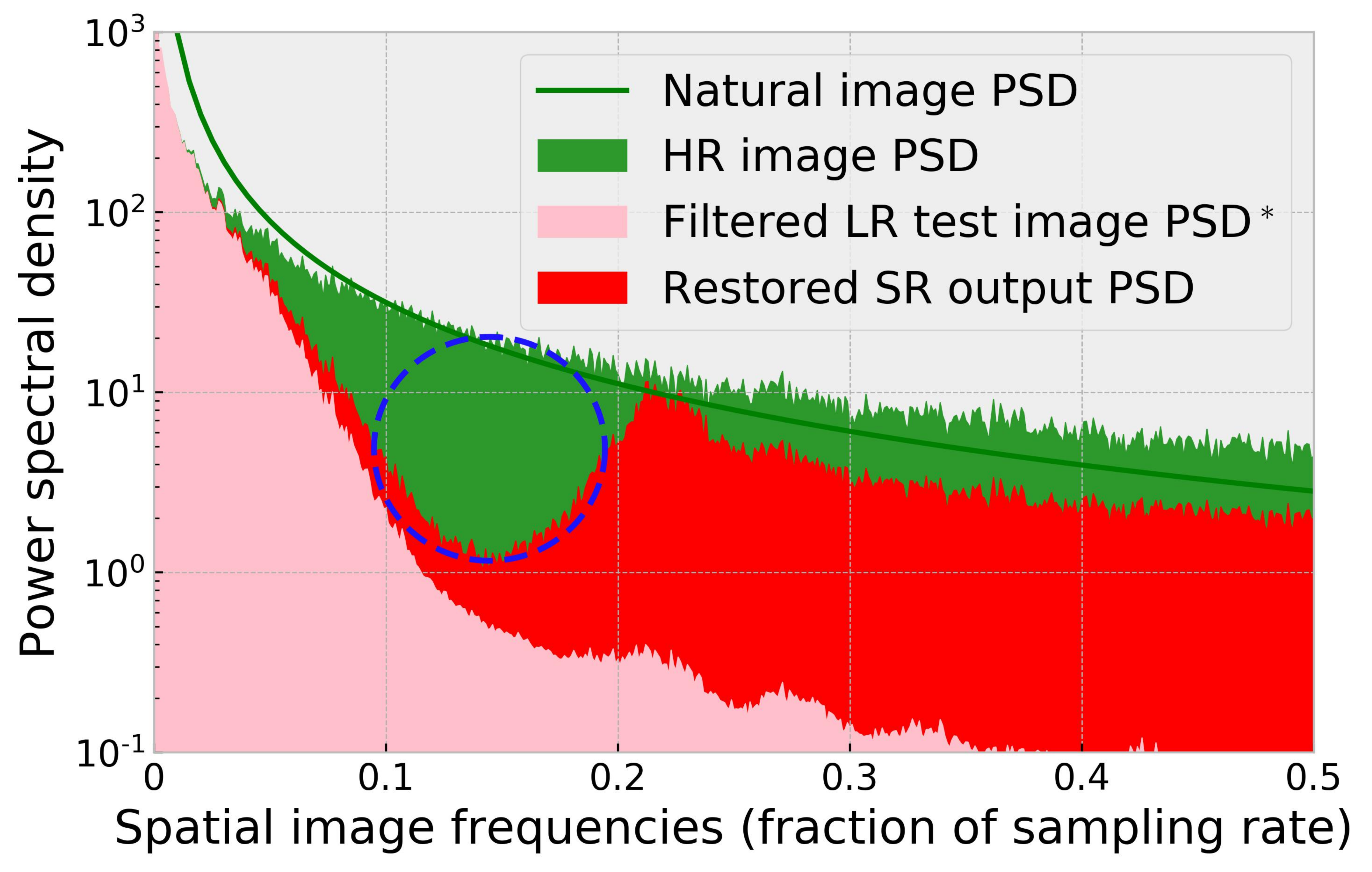}
		\label{fig:hole_without}}
	\subfigure[With SFM]{\includegraphics[width=.49\linewidth,trim={7 6 8 5},clip]{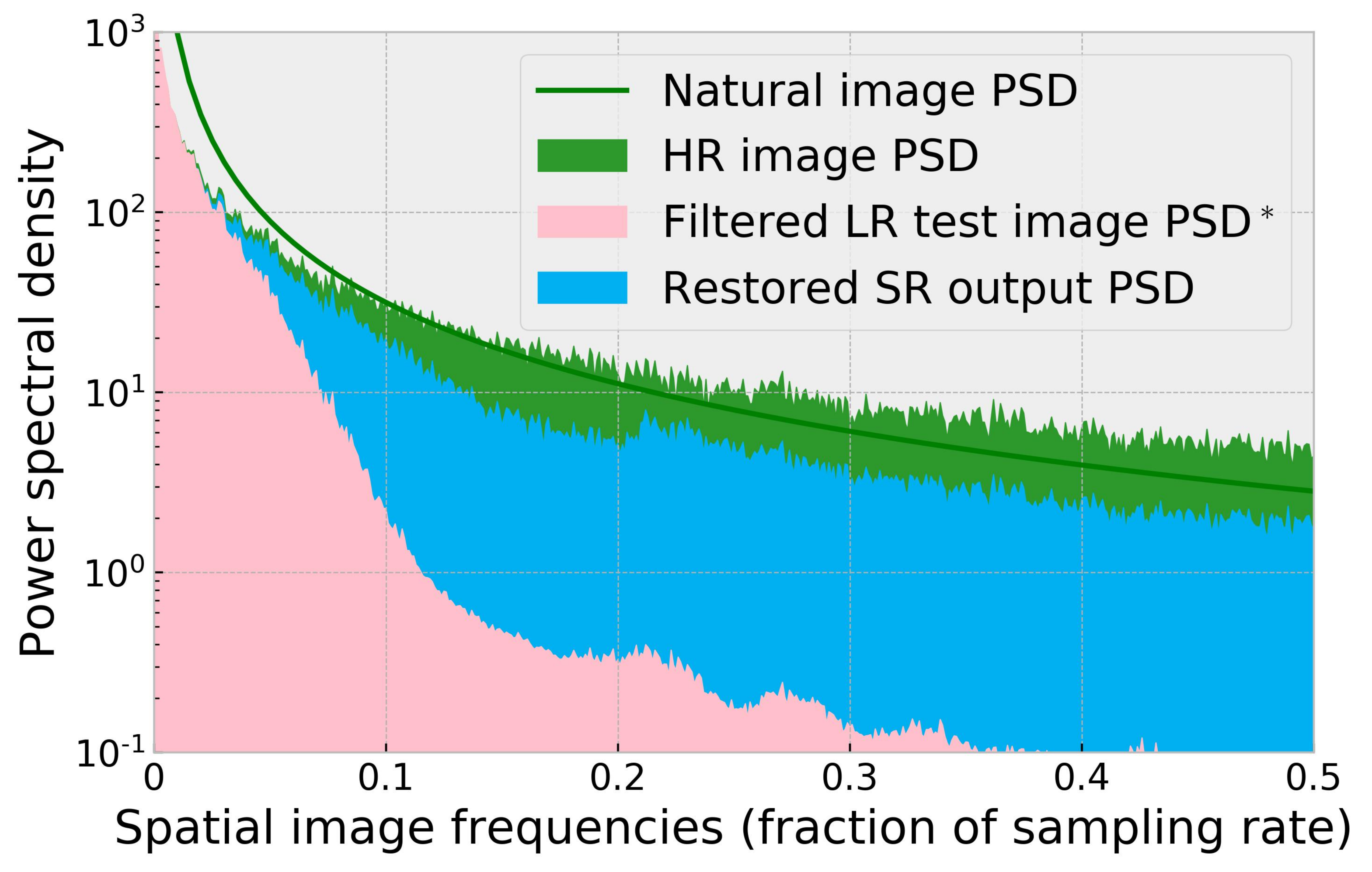}
		\label{fig:hole_with}}
	\caption{(a) Overview of our experimental setup, with image border colors corresponding to the plot colors shown in (b,c). We train 2 versions of the same network on the same degradation kernel ($F_1^{LP}$ anti-aliasing filter), one without and one with SFM, and test them using $F_2^{LP}$. (b) Average PSD (power spectral density) of HR images in green fill, with a green curve illustrating a typical natural-image PSD ($\alpha=1.5$~\cite{torralba2003statistics}). The pink fill illustrates the average PSD of the low-pass filtered LR test images (\textit{$^*$shown before downsampling for better visualization}). In red fill is the average PSD of the restored SR output image. The blue dashed circle highlights the learning gap due to degradation-kernel overfitting. (c) The same as (b), except that the output is that of the network trained with SFM. Results are averaged over 100 random samples.
	}
	\label{fig:hole_plot}
\end{figure}

\subsubsection{Frequency visualization of SR reconstructions} \label{subsubsec:SFM_SR_analysis}
SR networks tend to overfit to the blur kernels used in the degradation for obtaining the training images~\cite{SRMD}. To understand that phenomenon, we analyze in this section the relation between the frequency-domain effect of a blur kernel and the reconstruction of SR networks. We carry out the following experiment with a network trained with a unique and known blur kernel. We use the DIV2K~\cite{DIV2K} dataset to train a 20-block RRDB~\cite{RRDB} x$4$ SR network with images filtered by a Gaussian blur kernel called $F_1^{LP}$ (standard deviation $\sigma = 4.1$), shown in the top row of Fig.~\ref{fig:pipeline_hole}. We then run an inference on 100 test images filtered with a different Gaussian blur kernel called $F_2^{LP}$ ($\sigma = 7.4$), shown in the bottom row of Fig.~\ref{fig:pipeline_hole}, to analyze the potential network overfitting.

We present a frequency-domain visualization in Fig.~\ref{fig:hole_without}. The power spectral density (PSD) is the distribution of frequency content in an image. The typical PSD of an image (green curve) is modeled as $1/f^\alpha$, where $f$ is the spatial frequency, with $\alpha \in [1,2]$ and varying depending on the scene (natural vs. man-made)~\cite{burton1987color,field1987relations,tolhurst1992amplitude,torralba2003statistics}. The $1/f^\alpha$ trend is visible in the PSD of HR images (green fill). The degraded LR test images are obtained with a low-pass filter on the HR image, before downsampling, and their frequency components are mostly low frequencies (pink fill). The SR network outputs contain high-frequency components restored by the network (red fill). However, these frequencies are mainly above $0.2\pi$. This is only the range that was filtered out by the kernel used in creating the \textit{training} LR images. The low-pass kernel used in creating the test LR images filters out a larger range of frequencies; it has a lower cutoff than the training kernel (the reverse case is also problematic and is illustrated in \textit{Supplementary Material}). This causes a gap of missing frequency components not obtained in the restored SR output, illustrated with a blue dashed circle in Fig.~\ref{fig:hole_without}. The results suggest that an implicit conditional learning takes place in the SR network, on which we expand further in the following section.
The results of the network trained with 50\% SFM (masking applied to half of the training set) are shown in Fig.~\ref{fig:hole_with}. 
A key observation is that the missing frequency components are predicted to a far better extent when the network is trained with SFM.

\subsubsection{Implicit conditional learning} \label{subsubsec:SFM_SR_formulation}
As we explain in the Preliminaries of Sec.~\ref{subsubsec:SFM_SR_math}, the high-frequency components of the original HR images are removed by the anti-aliasing filter. If that filter is \textit{ideal}, it means that the low-frequency components are not affected and the high frequencies are perfectly removed. We propose that the SR networks in fact learn implicitly a conditional probability
\begin{equation}
P \left( I^{HR} \circledast F^{HP} \mid I^{HR} \circledast F^{LP} \right),
\end{equation}
where $F^{HP}$ and $F^{LP}$ are ideal high-pass and low-pass filters, applied to the high-resolution image $I^{HR}$, and $\circledast$ is the convolution operator. The low and high frequency ranges are theoretically defined as $[0,\pi/T]$ and $[\pi/T,\pi]$, which is the minimum condition (largest possible cutoff) to avoid aliasing for a downsampling rate $T$. The components of $I^{HR}$ that survive the low-pass filtering are the same frequencies contained in the LR image $I^{LR}$, when the filters $F$ are ideal. In other words, the frequency components of $I^{HR} \circledast F^{LP}$ are those remaining in the LR image that is the network input. 

The anti-aliasing filters are, in practice, not ideal, resulting in: (a) some low-frequency components of $I^{HR}$ being attenuated, (b) some high frequencies surviving the filtering and causing aliasing. Typically, the main issue is the first issue (a), because filters are chosen in a way to remove the visually-disturbing aliasing at the expense of attenuating some low frequencies. We expand further on this in \textit{Supplementary Material}, and derive that even with non-ideal filters, there is still conditional and residual learning components to predict a set of high-frequencies. These frequencies are, however, conditioned on a set of low-frequency components potentially attenuated by the non-ideal filter we call $F^{LP}_o$. This filter fully removes aliasing artifacts but can affect the low frequencies. The distribution can hence be defined by the components
\begin{equation}
P \left( I^{HR} \circledast F^{HP} \mid I^{HR} \circledast F^{LP}_o \right), \quad P \left( I^{HR} \circledast F^{LP} - I^{HR} \circledast F^{LP}_0 \mid I^{HR} \circledast F^{LP}_o \right).
\end{equation}
This is supported by our results in Fig.~\ref{fig:hole_plot}. The SR network trained with degradation kernel $F^{LP}_1$ ($\sigma=4.1$ in our experiment) restores the missing high frequencies of $I^{HR}$ that would be erased by $F^{LP}_1$. However, that is the case even though the test image is degraded by $F^{LP}_2 \neq F^{LP}_1$. As $F^{LP}_2$ ($\sigma=7.4$) removes a wider range of frequencies than $F^{LP}_1$, not predicted by the network, these frequencies remain missing. We observe a gap in the PSD of the output, highlighted by a blue dashed circle. This illustrates the degradation-kernel overfitting issue from a frequency-domain perspective. We also note that these missing frequency components are restored by the network trained with SFM.

\subsection{Extension to denoising} \label{subsec:SFM_denoising}
\begin{figure}[t]
	\centering
	\includegraphics[width=.49\linewidth,trim={5 2 0 5},clip]{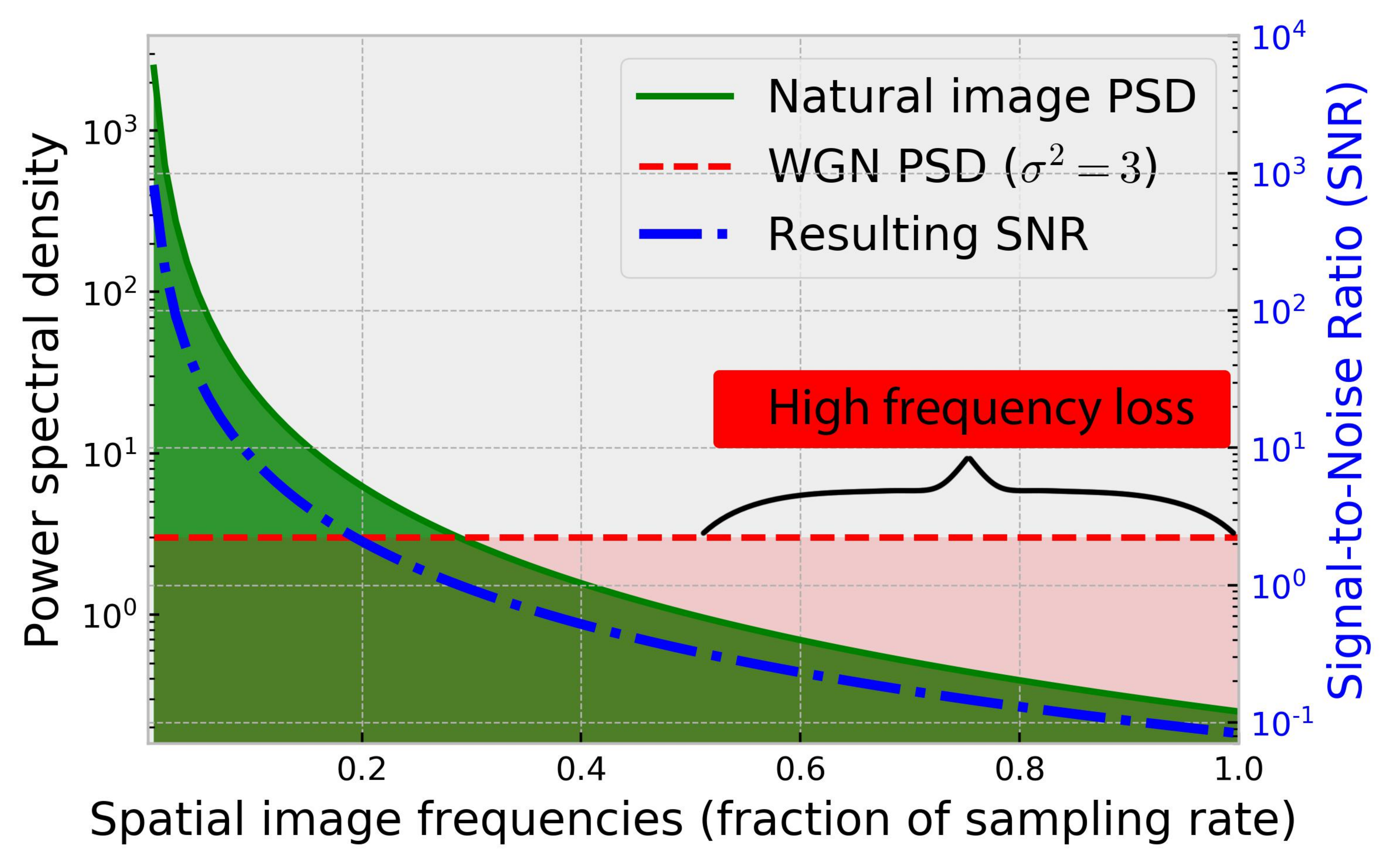}
	\includegraphics[width=.49\linewidth,trim={0 2 6 5},clip]{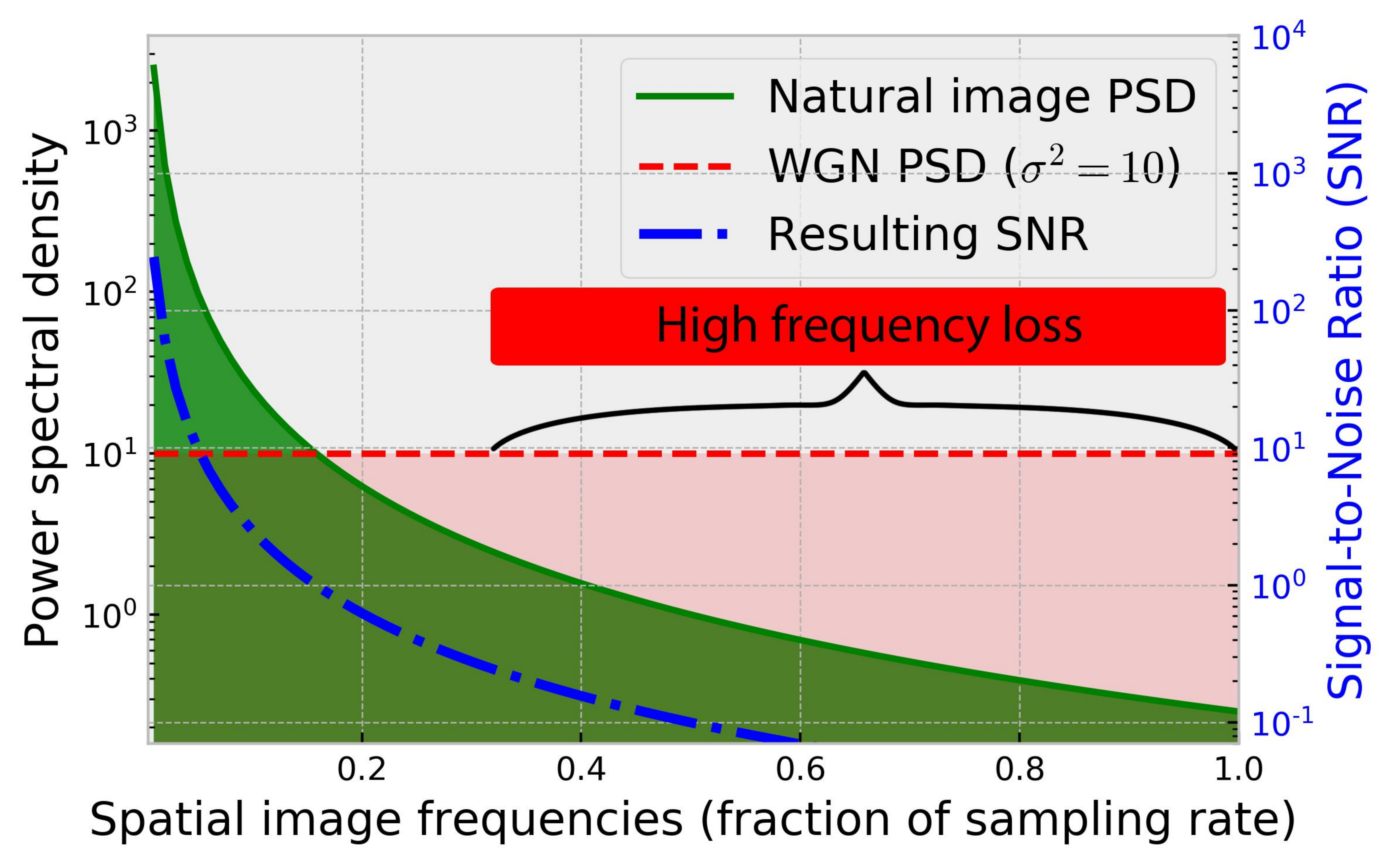}
	\caption{Natural image frequency content (image PSD) follows a power law as a function of spatial frequency. The plotted examples follow a power law with $\alpha = 2$~\cite{torralba2003statistics} and additive WGN ($\sigma^2=3$ on the left, and $\sigma^2=10$ on the right). The resulting SNR in the noisy image is exponentially smaller the higher the frequency, effectively causing a high frequency loss. The higher the noise level, the more frequency loss is incurred, and the more similar denoising becomes to our SR formulation.
	}
	\label{fig:PSD_plot}
\end{figure}

We highlight a connection between our conditional learning proposition and denoising. As discussed in Sec.~\ref{subsubsec:SFM_SR_analysis}, the average PSD of an image can be approximated by $1/f^\alpha$. The Gaussian noise samples added across pixels are independent and identically distributed. The PSD of the additive white Gaussian noise is uniform. 
Fig.~\ref{fig:PSD_plot} shows the PSD of a natural image following a power law with $\alpha = 2$, that of white Gaussian noise (WGN), and the resulting signal-to-noise ratio (SNR) when the WGN is added to the image. The resulting SNR decreases proportionally to $1/f^\alpha$.

The relation between SNR and frequency shows that with increasing frequency, the SNR becomes exponentially small. In other words, high frequencies are almost completely overtaken by the noise, while low frequencies are much less affected by it. And, \textit{the higher the noise level, the lower the starting frequency beyond which the SNR is significantly small}, as illustrated by Fig.~\ref{fig:PSD_plot}. This draws a direct connection to our SR analysis. Indeed, in both applications there exists an implicit conditional learning to predict lost high-frequency components given low-frequency ones that are less affected.

\section{Stochastic Frequency Masking (SFM)} \label{sec:SFM}

\subsection{Motivation and implementation} \label{subsec:SFM_implementation}
The objective of SFM is to improve the networks' prediction of high frequencies given lower ones, whether for SR or denoising. We achieve this by stochastically masking high-frequency bands from some of the training images in the learning phase, to encourage the conditional learning of the network.
Our masking is carried out by transforming an image to the frequency domain using the Discrete Cosine Transform (DCT) type II~\cite{ahmed1974discrete,strang1999discrete}, multiplying channel-wise by our stochastic mask, and lastly transforming the image back (Fig.~\ref{fig:teaser}). 
See \textit{Supplementary Material} for the implementation details of the DCT type we use.
We define frequency bands in the DCT domain over quarter-annulus areas, to cluster together similar-magnitude frequency content. Therefore, the SFM mask is delimited with a quarter-annulus area by setting the values of its inner and outer radii. We define two masking modes, the \textit{central mode} and the \textit{targeted mode}.

In the \textit{central mode}, the inner and outer radius limits $r_I$ and $r_O$ of the quarter-annulus are selected uniformly at random from $[0,r_M]$, where $r_M=\sqrt{a^2+b^2}$ is the maximum radius, with $(a,b)$ being the dimensions of the image. We ensure that $r_I < r_O$ by permuting the values if $r_I > r_O$. With this mode, the resulting probability of a given frequency band $r_{\omega}$ to be masked is
\begin{equation}
P(r_{\omega}=0) = P(r_I<r_{\omega}<r_O) = 2\left( \frac{r_{\omega}}{r_M} - \left(\frac{r_{\omega}}{r_M}\right)^2\right),
\end{equation}
which means the central bands are the more likely ones to be masked, with the likelihood \textit{slowly} decreasing the higher or the lower the frequencies are.
In the \textit{targeted mode}, a target frequency $r_C$ is selected along with a parameter $\sigma_{\Delta}$. The quarter-annulus is then delimited by $[r_C-\delta_I, r_C+\delta_O]$, where $\delta_I$ and $\delta_O$ are independently sampled from the half-normal $\Delta$ distribution $f_\Delta(\delta) = \sqrt{2}/\sqrt{\pi\sigma_\Delta^2} e^{-\delta^2/(2\sigma_\Delta^2)}$, $\forall \delta \geq 0$. Therefore, with this mode, the frequency $r_C$ is always masked, and the frequencies away from $r_C$ are less and less likely to be masked, with a normal distribution decay.

We use the \textit{central mode} for SR networks, and the \textit{targeted mode} with a high target $r_C$ for denoisers (Fig.~\ref{fig:teaser}). The former has a slow concave probability decay that allows to cover wider bands, while the latter has an exponential decay adapted for targeting very specific narrow bands. In both settings, the highest frequencies are most likely masked, and lower ones are masked with decaying probability. The \textit{central mode} indeed masks the highest frequencies in SR, because the central-band frequencies are the highest ones remaining \textit{in the HR image} after the anti-aliasing filter is applied. It is also worth noting for SR that SFM actually simulates the effect of different blur kernels by stochastically masking different frequency bands. 

\subsection{Learning SR with SFM} \label{subsec:SFM_SR_pipeline}
We apply SFM \textit{only} on the input training data. For the simulated-degradation datasets, SFM is applied in the process of generating the LR inputs. Namely, we apply SFM on the HR images before applying degradation model to generate the LR inputs (blur kernel convolution and downsampling). The target output of the network remains the original HR images. For the real datasets where the LR inputs are already given and the degradation model is unknown, we directly apply SFM on the LR inputs and keep the original HR images as ground-truth targets. Therefore, the networks trained with SFM do not use any additional data relative to those trained without SFM.

We apply \textit{the same} SFM settings for all deep learning experiments. During training, we apply SFM on 50\% of the training images, using the \textit{central mode} of SFM, as presented in Sec.~\ref{subsec:SFM_implementation}. Ablation studies with other rates are in our \textit{Supplementary Material}. We add SFM to the training of the original methods with no other modification.

\newcommand{\divk}[2]{ \subfigure[{#1}]{\includegraphics[width=0.24\linewidth,]{#2}}}
\begin{figure*}[t]
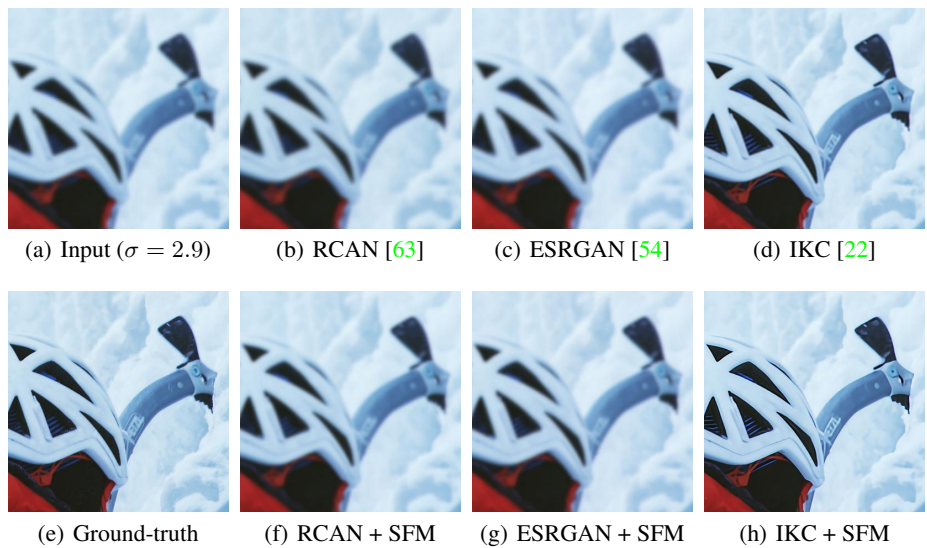

	\centering
	\divk{Input ($\sigma=2.9$)}{DIV2K/0844/LR}
	\divk{RCAN~\cite{RCAN}}{DIV2K/0844/RCAN}
	\divk{ESRGAN~\cite{RRDB}}{DIV2K/0844/ESRGAN}
	\divk{IKC~\cite{IKC}}{DIV2K/0844/IKC} \\
	\divk{Ground-truth}{DIV2K/0844/HR}
	\divk{RCAN + SFM}{DIV2K/0844/RCAN_SFM}
	\divk{ESRGAN + SFM}{DIV2K/0844/ESRGAN_SFM}
	\divk{IKC + SFM}{DIV2K/0844/IKC_SFM}
	\caption{Cropped SR results (x$4$ upscaling) with different methods (top row), and with the same methods trained with our SFM (bottom row), for image 0844 of the DIV2K benchmark. The visual quality of the results improves for all methods when trained with SFM (images best viewed on screen).}
	\label{fig:div2k_0844}
\end{figure*}

\subsection{Learning denoising with SFM} \label{subsec:experiments_den_pipeline}
We incorporate our SFM method into the denoiser training pipeline as follows. When training for additive white Gaussian noise (AWGN) removal, we apply SFM on the clean image before the synthetic noise is added. When the training images are real and the noise cannot be separated from the signal, we apply SFM on both the image and its noise. Hence, we ensure that networks trained with SFM do not utilize any additional training data relative to the baseline networks.
 
In all denoising experiments, and for all of the compared methods, we use \textit{the same} SFM settings. We again apply SFM on 50\% of the training images, and use the \textit{targeted mode} of our SFM (ablation studies including other rates are in our \textit{Supplementary Material}). We use a central band $r_C=0.85\,r_M$ and $\sigma_\Delta=0.15\,r_M$. As presented in Sec.~\ref{subsec:SFM_implementation}, this means that the highest frequency bands are masked with high likelihood, and lower frequencies are exponentially less likely to be masked the smaller they are. We add SFM to the training of the original methods with no other modification.

\section{Experiments} \label{sec:experiments}
We conduct experiments on state-of-the-art networks for various SR and denoising tasks. \textit{Supplementary Material} includes quantitative evaluations on additional SR upscaling factors, more visual results, a frequency assessment of our different results, and various ablation studies on different frequency band masking and varying SFM rates.

\subsection{SR: bicubic and Gaussian degradations}\label{subsubsec:experiments_SR_bicGauss}
\begin{table*}[t!]
\centering
\begin{tabular}{lc *{9}{c}}
\toprule
 & \multicolumn{10}{c}{{Test blur kernel ($g_{\sigma}$ is a Gaussian kernel, standard deviation $\sigma$)}} \\ \cline{2-11}

 & bicubic & $g_{1.7}$ & $g_{2.3}$ & $g_{2.9}$ & $g_{3.5}$ & $g_{4.1}$ & $g_{4.7}$ & $g_{5.3}$ & $g_{5.9}$ & $g_{6.5}$\\  \hhline{*{11}{-}}

RCAN~\cite{RCAN} & \cellcolor{lightgray} 29.18 & 23.80 & 24.08 & 23.76 & 23.35 & 22.98 & 22.38 & 22.16 & 21.86 & 21.72\\
RCAN with 50\% SFM & \cellcolor{lightgray} \textbf{29.32} & \textbf{24.21} & \textbf{24.64} & \textbf{24.19} & \textbf{23.72} & \textbf{23.27} & \textbf{22.54} & \textbf{22.23} & \textbf{21.91} & \textbf{21.79}\\
\hdashline

IKC~\cite{IKC} & \textbf{27.81} & \cellcolor{lightgray} 26.07 & \cellcolor{lightgray} 26.15 & \cellcolor{lightgray} 25.48 & \cellcolor{lightgray} 25.03 & 24.41 & 23.39 & 22.78 & 22.41 & 22.08\\
IKC with 50\% SFM & 27.78 & \cellcolor{lightgray} \textbf{26.09} & \cellcolor{lightgray} \textbf{26.18} & \cellcolor{lightgray} \textbf{25.52} & \cellcolor{lightgray} \textbf{25.11} & \textbf{24.52} & \textbf{23.54} & \textbf{22.97} & \textbf{22.62} & \textbf{22.35}\\
\hdashline

RRDB~\cite{RRDB} & \cellcolor{lightgray} 28.79 & 23.66& 23.72 & 23.68 & 23.29 & 22.75 & 22.32 & 22.08 & 21.83 & 21.40\\
RRDB with 50\% SFM & \cellcolor{lightgray} \textbf{29.10} & \textbf{23.81} & \textbf{23.99} & \textbf{23.79} & \textbf{23.41} & \textbf{22.90} & \textbf{22.53} & \textbf{22.37} & \textbf{21.98} & \textbf{21.56}\\
\hdashline

ESRGAN~\cite{RRDB} & \cellcolor{lightgray} 25.43 & 21.22 & 22.49 & 22.03 & 21.87 & 21.63 & 21.21 & 20.99 & 20.05 & 19.42\\
ESRGAN with 50\% SFM & \cellcolor{lightgray} \textbf{25.50} & \textbf{21.37} & \textbf{22.78} & 
\textbf{22.26} & \textbf{22.08} & \textbf{21.80} & 
\textbf{21.33} & \textbf{21.10} & \textbf{20.13} & \textbf{19.77}\\

\bottomrule

\end{tabular}
\caption{Single-image SR, with x$4$ upscaling factor, PSNR ($dB$) results on the DIV2K validation set. RCAN, RRDB and ESRGAN are trained using bicubic degradation, and IKC using Gaussian kernels ($\sigma \in [2.0, 4.0]$). Kernels seen in training are shaded in gray. The training setups of the different networks are presented in Sec.~\ref{subsubsec:experiments_SR_bicGauss}, and identical ones are used with SFM. We note that SFM improves the results of the various methods, even the IKC method that explicitly models kernels during its training improves by up to $0.27dB$ with SFM on unseen kernels.}
\vspace{-0.3cm}
\label{table:PSNR_DIV2K}
\end{table*}
\textbf{Methods}.
We evaluate our proposed SFM method on state-of-the-art SR networks that can be divided into 3 categories. In the first category, we evaluate RCAN~\cite{RCAN} and RRDB~\cite{RRDB}, which are networks that target pixel-wise distortion for a single degradation kernel. RCAN leverages a residual-in-residual structure and channel attention for efficient non-blind SR learning. RRDB~\cite{RRDB} employs a residual-in-residual dense block as its basic architecture unit. The second category covers perception-optimized methods for a single degradation kernel, and includes ESRGAN~\cite{RRDB}. It is a version of the RRDB network using a GAN for better SR perceptual quality and obtains the state-of-the-art results in this category. The last category includes algorithms for blind SR, we experiment on IKC~\cite{IKC}, which incorporates into the training of the SR network a blur-kernel estimation and modeling to explicitly address blind SR.
\textbf{Setup}.
We train all the models using the DIV2K~\cite{DIV2K} dataset, which is a high-quality dataset that is commonly used for single-image SR evaluation. RCAN, RRDB, and ESRGAN are trained with the bicubic degradation, and IKC with Gaussian kernels ($\sigma \in [0.2, 4.0]$~\cite{IKC}).
For all models, 16 LR patches of size $48 \times 48$ are extracted per training batch. All models are trained using the Adam optimizer~\cite{Adam} for 50 epochs. The initial learning rate is set to $10^{-4}$ and decreases by half every 10 epochs. Data augmentation is performed on the training images, which are randomly rotated by $90\degree$, $180\degree$, $270\degree$, and flipped horizontally.
\noindent \textbf{Results}. 
We evaluate the performance of the different methods on the image test set. To generate test LR images, we apply bicubic and Gaussian blur kernels on the DIV2K~\cite{DIV2K} validation set. We also evaluate all methods trained with 50\% SFM, following Sec.~\ref{subsec:SFM_SR_pipeline}. Table~\ref{table:PSNR_DIV2K} shows the PSNR results on x$4$ upscaling SR, with different blur kernels. Results show that the proposed SFM consistently improves the performance of the various SR networks on the different degradation kernels, even up to $0.27dB$ on an unseen test kernel for the recent IKC~\cite{IKC} that explicitly models kernels during training. We improve by up to $0.56dB$ for the other methods. 
With SFM, RRDB achieves comparable or better results than RCAN, which has double the number of parameters of RRDB. Sample visual results, without and with SFM training, are shown in Fig.~\ref{fig:div2k_0844}.

\subsection{SR: real-image degradations}
\begin{table}[t]
\centering
\begin{tabular}{l >{\centering}p{0.075\linewidth}>{\centering}p{0.075\linewidth}>{\centering}p{0.075\linewidth} | >{\centering}p{0.075\linewidth} >{\centering\arraybackslash}p{0.075\linewidth}}

\toprule
 & \multicolumn{5}{c}{Dataset and upscaling factor} \\ \cline{2-6}

& \multicolumn{3}{c}{RealSR~\cite{realsrdata}} & \multicolumn{2}{c}{SR-RAW~\cite{zoomindata}} \\ \cline{2-6}

Method & {x2} & {x3} & {x4} & {x4} & {x8} \\ \cline{1-6}

RCAN$^\ddagger$~\cite{RCAN} & 33.24 & 30.24 & 28.65 & 26.29 & 24.18\\
RCAN 50\% SFM & \textbf{33.32} & \textbf{30.29} & \textbf{28.75} & \textbf{26.42} & \textbf{24.50} \\
\hdashline
KMSR~\cite{KMSR} & 32.98 & 30.05 & 28.27 & 25.91 & 24.00\\
KMSR 50\% SFM  & \textbf{33.21} &  \textbf{30.11} &  \textbf{28.50} & \textbf{26.19} & \textbf{24.31} \\
\hdashline
IKC~\cite{IKC} & 33.07 & 30.03 & 28.29 & 25.87 & 24.19\\
IKC 50\% SFM & \textbf{33.12} & \textbf{30.25} & \textbf{28.42} & \textbf{25.93} & \textbf{24.25} \\

\bottomrule

\end{tabular}
\caption{PSNR ($dB$) results of blind image super-resolution on two real SR datasets, for the different available upscaling factors. $^\ddagger$RCAN is trained on the paired dataset collected from the same sensor as the testing dataset.}
\vspace{-0.5cm}
\label{table:PSNR_realSR}
\end{table}

\label{subsubsec:experiments_SR_real}
\textbf{Methods}. We train and evaluate the same SR models as the networks we use in Sec.~\ref{subsubsec:experiments_SR_bicGauss}, except for ESRGAN and RRDB, as ESRGAN is a perceptual-quality-driven method and does not achieve high PSNR, and RCAN outperforms RRDB according to our experiments in ~\ref{subsubsec:experiments_SR_bicGauss}. We also evaluate on KMSR~\cite{KMSR} for the real SR experiments. KMSR collects real blur kernels from real LR images to improve the generalization of the SR network on unseen kernels.
\noindent \textbf{Setup}. We train and evaluate the SR networks on two digital zoom datasets: the SR-RAW dataset~\cite{zoomindata} and the RealSR dataset~\cite{realsrdata}. The training setup of the SR networks is the same as in Sec.~\ref{subsubsec:experiments_SR_bicGauss}. Note that we follow the same training procedures for each method as in the original papers. IKC is trained with Gaussian kernels ($\sigma \in [0.2, 4.0]$) and KMSR with the blur kernels estimated from LR images in the dataset. RCAN is trained \textit{on the degradation of the test} data; a starting advantage over other methods.
\noindent \textbf{Results}. We evalute the SR methods on the corresponding datasets and present the results in Table~\ref{table:PSNR_realSR}. Each method is also trained with 50\% SFM, following Sec.~\ref{subsec:SFM_SR_pipeline}. SFM consistently improves all methods on all upscaling factors, pushing the state-of-the-art results by up to $0.23dB$ on both of these challenging real-image SR datasets.

\subsection{Denoising: AWGN} \label{subsubsec:experiments_den_AWGN}
\begin{table*}[t]
\centering
\begin{tabular}{lc *{9}{c}}
\toprule
& \multicolumn{10}{c}{{Test noise level (standard deviation of the stationary AWGN)}} \\ \hhline{~*{10}{-}}

& \cellcolor{lightgray} {10} & \cellcolor{lightgray} {20} & \cellcolor{lightgray} {30} & \cellcolor{lightgray} {40} & \cellcolor{lightgray} {50} & {60} & {70} & {80} & {90} & {100} \\  \cline{1-11}

DnCNN-B~\cite{dncnn} & 33.33 & 29.71 & 27.66 & 26.13 & 24.88 & 23.69 & 22.06 & 19.86 & 17.88 & 16.35 \\
DnCNN-B with 50\% SFM & \textbf{33.35} & \textbf{29.78} & \textbf{27.73} & \textbf{26.27} & \textbf{25.09} & \textbf{24.02} & \textbf{22.80} & \textbf{21.24} & \textbf{19.46} & \textbf{17.87} \\
\hdashline

Noise2Noise~\cite{lehtinen2018noise2noise} & \textbf{32.67} & 28.84 & 26.61 & 25.02 & 23.76 & 22.69 & 21.74 & 20.88 & 20.11 & 19.41 \\
Noise2Noise with 50\% SFM & 32.55 & \textbf{28.94} & \textbf{26.84} & \textbf{25.31} & \textbf{24.11} & \textbf{23.05} & \textbf{22.14} & \textbf{21.32} & \textbf{20.61} & \textbf{19.95} \\
\hdashline

Blind$^\ddagger$ N3Net~\cite{plotz2018neural} & \textbf{33.53} & \textbf{30.01} & \textbf{27.84} & 26.30 & 25.04 & 23.93 & 22.87 & 21.84 & 20.87 & 19.98 \\
N3Net with 50\% SFM & 33.41 & 29.86 & \textbf{27.84} & \textbf{26.38} & \textbf{25.19} & \textbf{24.15} & \textbf{23.20} & \textbf{22.32} & \textbf{21.51} & \textbf{20.78} \\
\hdashline

Blind$^\ddagger$ MemNet~\cite{tai2017memnet} & \textbf{33.51} & 29.75 & 27.61 & 26.06 & 24.87 & 23.83 & 22.67 & 21.00 & 18.92 & 17.16 \\
MemNet with 50\% SFM & 33.36 & \textbf{29.80} & \textbf{27.76} & \textbf{26.31} & \textbf{25.14} & \textbf{24.09} & \textbf{23.09} & \textbf{22.00} & \textbf{20.77} & \textbf{19.46} \\
\hdashline

RIDNet~\cite{anwar2019real} & \textbf{33.65} & \textbf{29.87} & 27.65 & 26.04 & 24.79 & 23.65 & 22.25 & 20.05 & 18.15 & 17.09 \\
RIDNet with 50\% SFM & 33.43 & 29.81 & \textbf{27.76} & \textbf{26.30} & \textbf{25.12} & \textbf{24.08} & \textbf{23.11} & \textbf{22.08} & \textbf{20.74} & \textbf{19.17} \\
\bottomrule

\end{tabular}
\caption{Blind AWGN removal PSNR ($dB$) results on the BSD68 set for different methods and noise levels. SFM improves the various methods, and the improvement increases with higher noise levels, supporting our hypothesis. We clamp test images to [0,255] as in camera pipelines. Denoisers are trained with levels up to $55$ (shaded in gray), thus half the test range is not seen in training. $^\ddagger$Re-trained under blind settings. 
}
\vspace{-0.5cm}
\label{table:PSNR_BSD68}
\end{table*}

\textbf{Methods}. We evaluate different state-of-the-art AWGN denoisers. 
DnCNN-B~\cite{dncnn} learns the noise residual rather than the final denoised image. Noise2Noise (N2N)~\cite{lehtinen2018noise2noise} learns only from noisy image pairs, with no ground-truth data. N3Net~\cite{plotz2018neural} relies on learning nearest neighbors similarity, to make use of different similar patches in an image for denoising. MemNet~\cite{tai2017memnet} follows residual learning with memory transition blocks. Lastly, RIDNet~\cite{anwar2019real} also does residual learning, but leverages feature attention blocks.
\noindent \textbf{Setup}. We train all methods on the 400 Berkeley images~\cite{martin2001database}, typically used to benchmark denoisers~\cite{chen2016trainable,schmidt2014shrinkage,dncnn}.  
All methods use the Adam optimizer with a starting learning rate of $10^{-3}$, except RIDNet that uses half that rate. We train for 50 epochs and synthesize noise instances per training batch. For blind denoising training, we follow the settings initially set in~\cite{dncnn}: noise is sampled from a Gaussian distribution with standard deviation chosen at random in $[0,55]$. This splits the range of test noise levels into levels seen or not seen during training, which provides further insights on generalization.
We also note that we use a U-Net~\cite{ronneberger2015u} for the architecture of N2N as in the original paper. For N2N, we apply SFM on top of the added noise, to preserve the particularity that N2N can be trained without ground-truth data.
\noindent \textbf{Results}. We evaluate all methods on the BSD68~\cite{roth2009fields} test set. Each method is also trained with 50\% SFM as explained in Sec.~\ref{subsec:experiments_den_pipeline} and the results are in Table~\ref{table:PSNR_BSD68}. SFM improves the performance of a variety of different state-of-the-art denoising methods on high noise levels (seen during training, such as $40$ and $50$, or not even seen), and the results support our hypothesis presented in Sec.~\ref{subsec:SFM_denoising} that \textit{the higher the noise level} the more similar is denoising to SR and the more applicable is SFM. Indeed, the higher the noise level the larger the improvement of SFM, and this trend is true across all methods. Fig.~\ref{fig:bsd68_image_67} presents sample visual results.

\newcommand{\bsd}[2]{ 
\subfigure[{#1}]{\includegraphics[width=0.15\linewidth,trim={80 30 80 30},clip]{#2}}}

\begin{figure*}[ht]
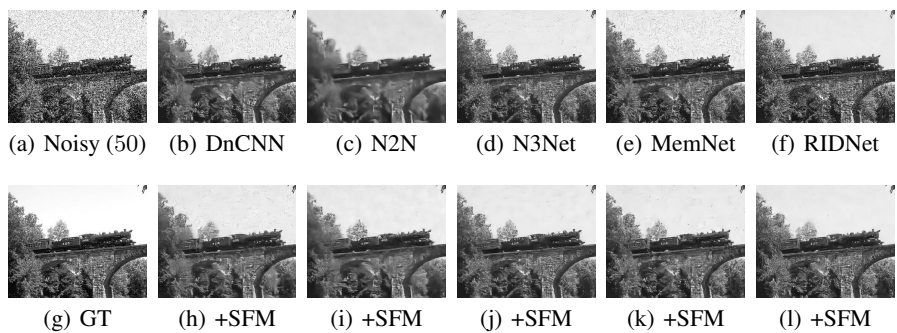

	\captionsetup[subfigure]{labelformat=empty}
	\centering
	\bsd{Noisy ($50$)}{bsd68/Noisy_Img_67_std50}
	\bsd{DnCNN}{bsd68/DnCNN_Img_67_std50}
	\bsd{N2N}{bsd68/N2N_Img_67_std50}
	\bsd{N3Net}{bsd68/N3Net_Img_67_std50}
	\bsd{MemNet}{bsd68/MemNet_Img_67_std50}
	\bsd{RIDNet}{bsd68/RIDNet_Img_67_std50} \\
	\bsd{GT}{bsd68/GT_Img_67}
	\bsd{+SFM}{bsd68/DnCNN_SDM_Img_67_std50}
	\bsd{+SFM}{bsd68/N2N_SDM_Img_67_std50}
	\bsd{+SFM}{bsd68/N3Net_SDM_Img_67_std50}
	\bsd{+SFM}{bsd68/MemNet_SDM_Img_67_std50}
	\bsd{+SFM}{bsd68/RIDNet_SDM_Img_67_std50}
	\caption{Denoising results with different methods (top row), and with the same method trained with our SFM (bottom row), for the last image (\#67) of the BSD68 benchmark.}
	\label{fig:bsd68_image_67}
\end{figure*}

\subsection{Denoising: real Poisson-Gaussian images} \label{subsubsec:experiments_den_Fluo}

\begin{table}[t]
	\centering
	\begin{tabular}{l >{\centering}p{0.070\linewidth}>{\centering}p{0.070\linewidth}>{\centering}p{0.070\linewidth}>{\centering}p{0.070\linewidth}>{\centering}p{0.070\linewidth}|>{\centering}p{0.070\linewidth}>{\centering}p{0.070\linewidth}>{\centering}p{0.070\linewidth}>{\centering}p{0.070\linewidth} >{\centering\arraybackslash}p{0.070\linewidth}}
		
		\toprule
		& \multicolumn{10}{c}{\# raw images for averaging} \\ \cline{2-11}
		
		& \multicolumn{5}{c|}{Mixed test set~\cite{zhang2019poisson}} & \multicolumn{5}{c}{Two-photon test set~\cite{zhang2019poisson}} \\ \cline{2-11}
		
		Method & {16} & {8} & {4} & {2} & {1} & {16} & {8} & {4} & {2} & {1} \\  \cline{2-11}
		PURE-LET~\cite{luisier2011image} & 39.59 & 37.25 & 35.29 & 33.49 & 31.95 & 37.06 & 34.66 & 33.50 & 32.61 & 31.89 \\
		VST+KSVD~\cite{aharon2006k} & 40.36 & 37.79 & 35.84 & 33.69 & 32.02 & 38.01 & 35.31 & 34.02 & 32.95 & 31.91 \\
		VST+WNNM~\cite{gu2014weighted} & 40.45 & 37.95 & 36.04 & 34.04 & 32.52 & 38.03 & 35.41 & 34.19 & 33.24 & 32.35 \\
		VST+BM3D~\cite{BM3D} & 40.61 & 38.01 & 36.05 & 34.09 & 32.71  & 38.24 & 35.49 & 34.25 & 33.33 & 32.48\\
		VST+EPLL~\cite{zoran2011learning} & 40.83 & 38.12 & 36.08 & 34.07 & 32.61 & 38.55 & 35.66 & 34.35 & 33.37 & 32.45 \\
		
		N2S~\cite{batson2019noise2self} & 36.67 & 35.47 & 34.66 & 33.15 & 31.87 & 34.88 & 33.48 & 32.66 & 31.81 & 30.51 \\
		N2S 50\% SFM & 36.60 & 35.62 & 34.59 & 33.44 & 32.40 & 34.39 & 33.14 & 32.48 & 31.84 & 30.92 \\
		
		N2N~\cite{lehtinen2018noise2noise} & 41.45 & 39.43 & 37.59 & 36.40 & 35.40 & 38.37 & 35.82 & 34.56 & 33.58 & 32.70 \\
		N2N 50\% SFM & \textbf{41.48} & \textbf{39.46} & \textbf{37.78} & \textbf{36.43} & \textbf{35.50} & \textbf{38.78} & \textbf{36.10} & \textbf{34.85} & \textbf{33.90} & \textbf{33.05} \\		
		
		\bottomrule
		
	\end{tabular}
	\caption{PSNR ($dB$) denoising results on fluorescence microscopy images with Poisson-Gaussian noise. We train under blind settings and apply SFM on noisy input images to preserve the fact that N2S and N2N can be trained without clean images.}
	\vspace{-0.3cm}
	\label{table:PSNR_fluo}
\end{table}

\textbf{Methods}. Classical methods are often a good choice for denoising in the absence of ground-truth datasets. PURE-LET~\cite{luisier2011image} is specifically aimed at Poisson-Gaussian denoising, and KSVD~\cite{aharon2006k}, WNNM~\cite{gu2014weighted}, BM3D~\cite{BM3D}, and EPLL~\cite{zoran2011learning} are designed for Gaussian denoising. Recently, learning methods were presented such as N2S~\cite{batson2019noise2self} (and the similar, but less general, N2V~\cite{krull2019noise2void}) that can learn from a dataset of only noisy images, and N2N~\cite{lehtinen2018noise2noise} that can learn from a dataset of only noisy image pairs. We incorporate SFM into the learning-based methods.
\noindent \textbf{Setup}. We train the learning-based methods on the recent real fluorescence microscopy dataset~\cite{zhang2019poisson}. The noise in that dataset follows a Poisson-Gaussian distribution, and the image registration is of high quality due to the stability of the microscopes, thus yielding reliable ground-truth, obtained by averaging 50 repeated captures. The noise parameters are estimated using the fitting approach in~\cite{foi2008practical} for all classical denoisers as they are not blind. Additionally, the parameters are used for the variance-stabilization transform (VST)~\cite{makitalo2012optimal} needed to transform into a Gaussian-denoising problem for the Gaussian-oriented methods. In contrast, the learning methods can directly be applied under blind settings. We train N2S/N2N using a U-Net~\cite{ronneberger2015u} architecture, for 100/400 epochs using the Adam optimizer with a starting learning rate of $10^{-5}$/$10^{-4}$~\cite{zhang2019poisson}. 
\noindent \textbf{Results}. We evaluate all methods on the mixed and two-photon microscopy test sets~\cite{zhang2019poisson}. We also train the learning methods with 50\% SFM as explained in Sec.~\ref{subsec:experiments_den_pipeline}, and present the results in Table~\ref{table:PSNR_fluo} (visual results in Fig.~\ref{fig:fluo_image_crops}). A larger number of raw images used for averaging is equivalent to a lower noise level. We observe that N2N with SFM achieves the state-of-the-art performance on both benchmarks and for all noise levels, with an improvement of up to $0.42dB$. We also note that the improvements of SFM are larger on the more challenging two-photon test set where the noise levels are higher on average. SFM does not consistently improve N2S, however, this is expected. In fact, unlike other methods, N2S trains to predict a subset of an image given a surrounding subset. It applies spatial masking where the mask is made up of random pixels that interferes with the frequency components. For these reasons, N2S is not very compatible with SFM, which nonetheless improves the N2S results on the largest noise levels in both test sets.

\newcommand{\fluocroptop}[1]{ \subfigure{\includegraphics[width=0.156\linewidth,trim={80 0 80 0},clip]{#1}}}
\newcommand{\fluocrop}[2]{ 
\subfigure[{#1}]{\includegraphics[width=0.156\linewidth,trim={80 0 80 0},clip]{#2}}}
\begin{figure}[h]
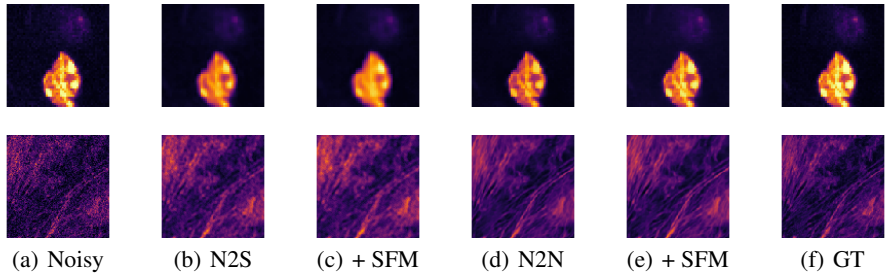

	\centering
	\fluocroptop{fluo_crops/MICE_16_1_noisy}
	\fluocroptop{fluo_crops/MICE_16_1_denoised0n2s}
	\fluocroptop{fluo_crops/MICE_16_1_denoised05n2s}
	\fluocroptop{fluo_crops/MICE_16_1_denoised0n2n}
	\fluocroptop{fluo_crops/MICE_16_1_denoised05n2n}
	\fluocroptop{fluo_crops/MICE_16_1_gt} \\
	
	\setcounter{subfigure}{0}
	\fluocrop{Noisy}{fluo_crops/BPAE_G_1_1_noisy}
	\fluocrop{N2S}{fluo_crops/BPAE_G_1_1_denoised0n2s}
	\fluocrop{+ SFM}{fluo_crops/BPAE_G_1_1_denoised05n2s}
	\fluocrop{N2N}{fluo_crops/BPAE_G_1_1_denoised0n2n}
	\fluocrop{+ SFM}{fluo_crops/BPAE_G_1_1_denoised05n2n}
	\fluocrop{GT}{fluo_crops/BPAE_G_1_1_gt}
	
	\caption{Cropped sample results for denoising image (a) from the real fluorescence microscopy denoising dataset. The top row averages 16 raw images (MICE scan) to obtain (a), and the bottom row directly denoises from 1 image only (BPAE scan). The `ground-truth' image (f) is estimated by averaging 50 raw images~\cite{zhang2019poisson}.}
	\label{fig:fluo_image_crops}
\end{figure}

\section{Conclusion} \label{sec:conclusion}
We analyze the degradation-kernel overfitting of SR networks in the frequency domain. Our frequency-domain insights reveal an implicit conditional learning that also extends to denoising, especially on high noise levels. Building on our analysis, we present SFM, a technique to improve SR and denoising networks, without increasing the size of the training set or any cost at test time.
We conduct extensive experiments on state-of-the-art networks for both restoration tasks. We evaluate SR with synthetic degradations, real-image SR, Gaussian denoising and real-image Poisson-Gaussian denoising, showing improved performance, notably on generalization, when using SFM.

\clearpage
\bibliographystyle{splncs04}
\bibliography{egbib}
\end{document}